\documentclass[preprint]{aastex}

\usepackage{amsmath}	
\usepackage{amssymb}	
\usepackage{graphicx,color}
\usepackage{natbib}

\shorttitle{Thermal conduction as a mechanism to make the heliosheath thinner}
\shortauthors{Izmodenov et al.}

\begin{document}

\title{ELECTRON THERMAL CONDUCTION AS A POSSIBLE PHYSICAL MECHANISM TO MAKE THE INNER HELIOSHEATH THINNER}

\author{V. V. Izmodenov \altaffilmark{1,2,3} \and D. B. Alexashov \altaffilmark{2,3} \and M. S. Ruderman \altaffilmark{2,4}}

\altaffiltext{1}{Lomonosov Moscow State University; izmod@ipmnet.ru}
\altaffiltext{2}{Space Research Institute (IKI) of Russian Academy of Sciences}
\altaffiltext{3}{Institute for Problems in Mechanics Russian Academy of Sciences}
\altaffiltext{4}{School of Mathematics and Statistics, University of Sheffield, Hicks Building, Sheffield S3 7RH, UK}

\begin{abstract}

We show that the electron thermal conductivity may strongly affect the heliosheath plasma flow and the global pattern of the solar wind (SW) interaction with the local interstellar medium (LISM). In particular, it leads to strong reduction of the inner heliosheath thickness that makes possible to explain (qualitatively) why Voyager~1 (V1) has crossed the heliopause at unexpectedly small heliocentric distance of 122 AU.
To estimate the effect of thermal conductivity we  consider a limiting case when thermal conduction is very effective. To do that we assume the plasma flow in the entire heliosphere is nearly isothermal.

Due to this effect, the heliospheric distance of the termination shock has increased by about 15~AU in V1 direction compared to the adiabatic case with $\gamma = 5/3$. The heliospheric distance of the heliopause has decreased by about 27~AU. As a result, the thickness of the inner heliosheath in the model has decreased by about 42~AU and become equal to 32~AU.
\end{abstract}


\section{INTRODUCTION}
\label{intro}

\cite{Stone2013} and \cite{Krimigis2013}  have reported a sudden drop of the fluxes of the heliospheric energetic particles and a substantial increase of the galactic cosmic ray (GCR) fluxes in August 2013 at the heliocentric distance of $\sim$122 AU.
Such a behavior of the energetic particle fluxes could be treated as crossing of the heliopause by Voyager~1 (V1). Such a conclusion has not been made since the magnetic field did not change its direction \citep{Burlaga2013} as it could be expected since it is very improbable that the direction of the interstellar magnetic field (IsMF) is the same as that of the heliospheric magnetic field (HMF).

\cite{Gurnett2013} analyzed the V1 kHz emission event occurred in April 2013, and showed that the registered radio-frequency gives the estimate for the plasma number density of 0.08~cm$^{-3}$\/. This estimate is substantially larger than the solar wind number density and corresponds to the expected density in the interstellar medium. Therefore, \cite{Gurnett2013} concluded that V1 was inside  LISM in 2013 and crossed the heliopause in August 2012.

The V1 crossing of the heliopause at 122 AU has not been expected by a part of the heliospheric community since the global models of the SW/LISM interaction suggest that the thickness of the inner heliosheath in the V1 direction should be of the order of 50-70 AU depending on the model \citep[see, e.g.][]{Izmodenov2013}. Several ideas to resolve this problem appeared recently in the literature. \cite{Borovikov2014} suggested that the smaller distance is connected with the instabilities of the heliopause. \cite{Schwadron2013} argued for the interstellar flux transfer effect. \cite{Swisdak2013} and \cite{Opher2013} explained the observed behavior by magnetic reconnection in the inner heliosheath and at the heliopause.

Here it is worth noting that there are theories suggesting that V1 is still inside the heliosphere \citep{Bar2013,Fisk2014}. For the rest of this paper we assume that the crossing occurred at 122 AU and  estimate how the effect of thermal conduction influences the plasma flow in the inner heliosheath and the positions of the termination shock and heliopause. The importance of the electron thermal conduction in the heliosheath plasma flow has been recently pointed out by \cite{Baranov2013}. They have shown that it is the most important dissipative process. The relative importance of thermal conduction is characterized by the Peclet number:  $Pe = k_B n_e L V/k$, where $k_B$ is the Boltzman constant, $n_e$ is the electron number density, $L$ is the characteristic spatial scale of the problem, $V$ is the characteristic plasma speed, $k$ is the coefficient of the parallel thermal conduction.

When ${\rm Pe} \gg 1$ the effect of thermal conduction can be neglected, while it is very important when ${\rm Pe} \lesssim 1$. \cite{Baranov2013} have estimated that ${\rm Pe} \approx 5$ in the outer heliosheath and ${\rm Pe} \lesssim 1$ in the inner heliosheath. Hence, we can expect that the effect of thermal conduction is more pronounced in the inner heliosheath than in the outer one.

\section{PROBLEM FORMULATION, NUMERICAL METHOD, AND BOUNDARY CONDITIONS}\label{gov_eqns}

We consider the interaction of the solar wind with the local interstellar medium using kinetic-hydrodynamic approach. In accordance with this approach the plasma component is described by the ideal magneto-hydrodynamic (MHD) equations, while the neutral component is described using the kinetic equation. The latter equation is solved using the Monte-Carlo method. To solve the ideal MHD equations we use finite-volume high-order Godunov scheme that includes 3D adaptive moving grid with discontinuities capturing and fitting capabilities, Harten-Lax-van Leer Discontinuity (HLLD) MHD Riemann solver and Chakravarthy-Osher TVD procedure. The detailed description of the equations and numerical method can be found, e.g.\ in \cite{Izmodenov2009}.

In our calculations we have used the following values for the parameters in the local interstellar medium (LISM) and solar wind. In LISM: The proton number density is $n_{\rm p,LISM} = 0.06\;\mbox{cm}^{-3}$\/, the H-atom number density $n_{\rm H,LISM} = 0.18\;\mbox{cm}^{-3}$\/, the temperature of both plasma and neutral component $T_{\rm LISM} = 6530\;\mbox{K}$, the velocity $V_{\rm LISM} = 26.4\;\mbox{km/s}$\/, the magnetic field magnitude $B_{\rm LISM} = 4.4\;\mu\mbox{G}$, and the angle between the velocity and magnetic field ${\rm angle}(BV) = 20^\circ$. In the solar wind we imposed the boundary conditions at 1~AU: spherical solar wind without magnetic field with  the proton number density $n_{\rm E} = 7.39\;\mbox{cm}^{-3}$\/, the radial velocity $V_{\rm E} = 432\;\mbox{km/s}$\/, and the plasma temperature $T_{\rm E} = 67800\;\mbox{K}$.

To consider the thermal conduction we need to include the corresponding term in the energy equation. To do this we need to modify substantially the Godunov method that we use. Another complication is related to the fact that, in the presence of magnetic field, the thermal conduction is strongly anisotropic. The heat flux is mainly directed along the magnetic field, while the heat flux perpendicular to the magnetic field is almost completely suppressed. To avoid the complications but still to estimate the effect of thermal conduction, one could consider a limiting case of isothermal flow in the entire heliosphere. The isothermal flow corresponds to the case of very strong thermal conduction when ${\rm Pe} \ll 1$. Mathematically the isothermal flows of perfect gases correspond to the flows with the polytropic index $\gamma = 1$. It follows from the Clapeyron equation that, in an adiabatic flow, the pressure $P$ and density $\rho$ are related by $P = \rho R T$, with the temperature $T = \mbox{const}$. The adiabatic motion of the plasma is described by the ideal MHD equations with the the pressure and density related by $P \propto \rho^\gamma$\/, where $\gamma$ is the adiabatic index. For fully ionized plasmas $\gamma = 5/3$. The adiabatic and isothermal flows are two limiting cases of negligible and very strong heat conduction. Comparing these two cases would give us an estimate of the possible thermal conduction effect on the plasma flow in the SW/LISM interaction region.

Since consideration of the isothermal flow requires significant modification of our numerical code, we consider instead nearly isothermal polytropic flow in the entire heliosphere (i.e.\ the region inside the heliopause) with $\gamma = 1.06$. The reason why we did not take $\gamma = 1$ is purely technical: Our numerical code does not allow $\gamma = 1$. The effect of thermal conduction in this case would be slightly smaller than in the exactly isothermal case, but it is sufficient enough to be useful for the goals of this paper. For the outer heliosheath (i.e.\ the region outside of the heliopause) we assume $\gamma = 5/3$

We should emphasis that, of course, we do not pretend that this simplified approach  describes the effect of thermal conduction correctly in details.  We only hope that our simple model gives a numerical estimate of how the thermal conduction influences the global size and shape of the heliospheric interface.

\section{NUMERICAL RESULTS}
\label{results}

\begin{figure}[ht!]
\centerline{}
\includegraphics[width=0.6\textwidth,angle=0]{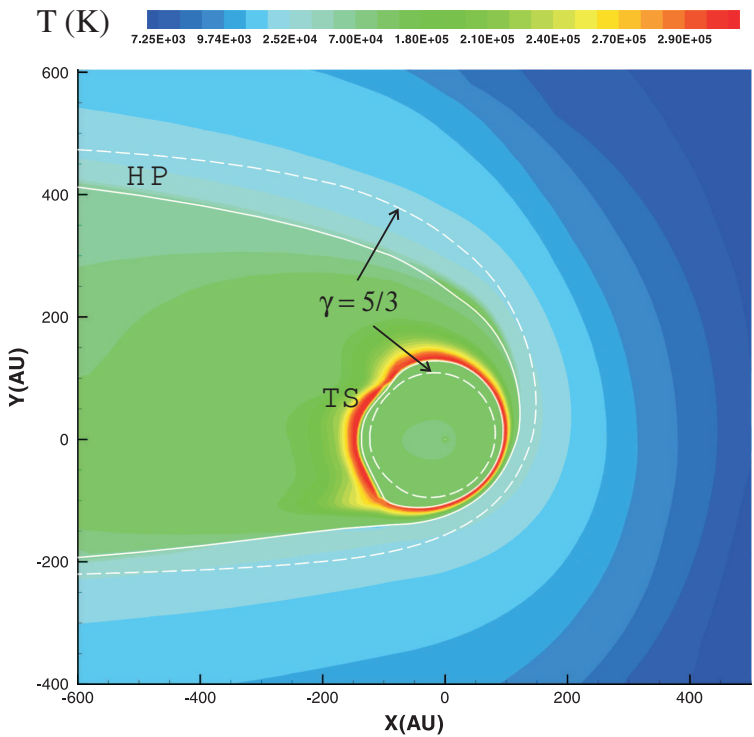}
\caption{The termination shock and heliopause for the models
with $\gamma = 5/3$ taken everywhere (dashed curves) and with $\gamma = 1.06$ in the inner 
heliosheath (solid curves).
Isolines show the plasma temperature calculated for the model with $\gamma = 1.06$.
}
\label{fig:temp}
\end{figure}

In this section we present the results of our numerical calculations.
Fig.~\ref{fig:temp} presents the locations of the termination shock and heliopause for the models with $\gamma = 1.06$ (solid curves) and $\gamma = 5/3$ (dashed curves) in the inner heliosheath. It is seen that the termination shock moves out and the heliopause moves in in the model with $\gamma = 1.06$ as compared to the other model. This is an expected effect of the thermal conduction that leads to a substantial reduction of the temperature and thermal pressure in the heliosheath (especially on the upwind side). Fig.~\ref{fig:temp} shows also the isolines of the plasma temperature for the model with $\gamma = 1.06$. It is seen from the figure that in the entire heliosphere the temperature is nearly constant ($\sim 1.2-3 \times 10^5$~K).

 It is interesting to note that despite $\gamma = 1.06$ is close to one and the supersonic solar wind is close to isothermal, there is a temperature jump at the TS.
This result directly follows from the Rankine-Hugoniot relations at the shock for the gas with $\gamma=1.06$.

\begin{figure}[ht!]
\centerline{}
\includegraphics[width=0.6\textwidth,angle=0]{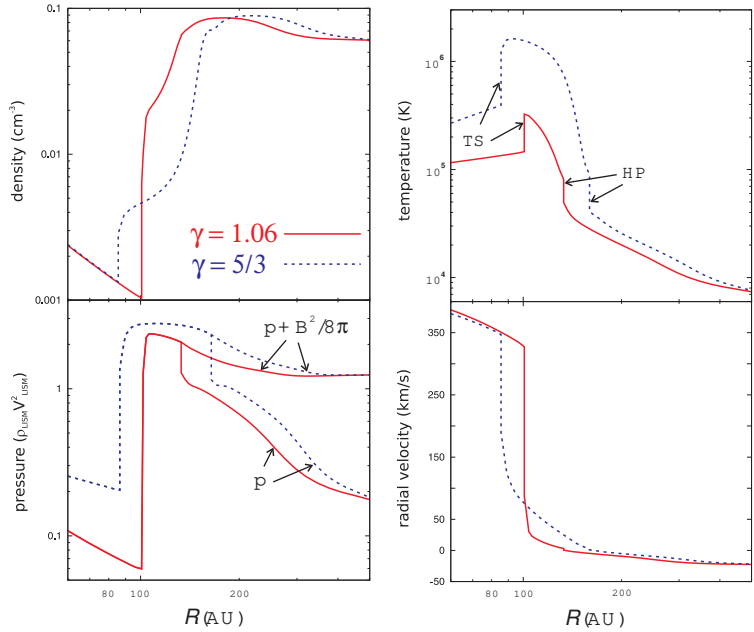}
\caption{The dependence of the plasma density, pressure, total pressure, temperature and radial 
velocity on the heliospheric distance in the V1 direction. The black lines correspond to the 
case where $\gamma = 5/3$ was taken everywhere, while the red lines to the case where $\gamma = 
1.06$ in the inner heliosheath.}
\label{fig:n_p_t_v}
\end{figure}

Figure~\ref{fig:n_p_t_v} displays the dependence of the plasma density, pressure, total pressure (plasma plus magnetic), temperature and radial velocity on the heliospheric distance in the V1 direction. The blue dashed lines correspond to the case where $\gamma = 5/3$ was taken everywhere, while the red solid lines to the case where $\gamma = 1.06$ inside the heliopause. This figure allows us to determine directly that the distance from the Sun to the termination shock increases by about 15~AU in the V1 direction, while the distance from the Sun to the heliopause decreases by about 27~AU. As a result, the thickness of the inner heliosheath decreases by about 42~AU and becomes only 32~AU.

\section{SUMMARY AND CONCLUSIONS}
\label{sum}

In this article, we have studied the effect of thermal conduction in the inner heliosheath on the plasma flow and the positions of the termination shock and heliopause. To simplify the problem, instead of taking into account the term describing the thermal conduction in the energy equation, we have modelled nearly isothermal flow by reducing polytropic index  to $\gamma = 1.06$ when describing the flow of the heliospheric plasma inside the heliopause. The main effect of this reduction is the following. The heliospheric distance of the termination shock has increased by about 15~AU in comparison with the case where $\gamma = 5/3$ everywhere. The heliospheric distance of the heliopause has decreased by about 27~AU. As a result, the thickness of the inner heliosheath has decreased by about 42~AU and become equal to 32~AU.

Our simplified model demonstrates that the account of thermal conduction strongly influences the plasma flow in the SW/LISM interaction region and the positions of the TS and HP. Therefore, more complex models including anisotropic thermal conduction effects  are needed to analyze data obtained by Voyager~1 and 2 and IBEX spacecraft, and to understand SW/LISM interaction, and to investigate the role of electrons as a separate fluid in this context, as already considered in a paper by \cite{ChalovFahr2013}. 

\section*{Acknowledgements}
Heliospheric part of the research has been partly supported by RFBR grant 13-01-00265. Numerical modeling of nearly isothermal stellar wind flow has been done in the frame of RNF grant 14-12-01096. MSR acknowledges the support by the STFC grant.

\end{document}